\documentclass{appolb}
\usepackage{graphicx}
\usepackage{amssymb}   
\def\Pom{{\bf I\!P}}

\begin{document}

\title{ PION-PION SCATTERING ABOVE RESONANCES }
   \author{A. Szczurek$^{1},^{2}$  \\
   {\it $^{1}$ Institute of Nuclear Physics, PL-31-342 Cracow, Poland  } \\
   {\it $^{2}$ University of Rzesz\'ow, PL-35-959 Rzesz\'ow, Poland } \\ }
\maketitle

\begin{abstract}

We model large angle intermediate energy pion-pion
scattering by the pQCD two-gluon (2g) exchange contribution
and discuss the onset of the dominance of
the Glauber-Gribov-Landshoff (GGL) component.
The  pQCD 2g exchange contribution becomes substantial already at
$|t|\sim$ 1 GeV$^{2}$, but the pQCD exchange dominance is
deferred to $|t|\sim$ 3 GeV$^{2}$ because of competing
multiple soft rescattering effects. Based on the $NN$ and $\pi N$
total cross section data and Regge factorization, we evaluate
the dominant soft contribution to the $\pi \pi$ total cross section
and find the results consistent with the ones deduced earlier from the
absorption model analysis of the $\pi N \to X N, X \Delta$
data.
\end{abstract}


\section{Introduction}

The pion-pion scattering, although not directly accessible experimentally,
is of special theoretical interest. At low energy it is the fundamental
testing ground of chiral perturbation theory
\cite{Weinberg,ChPT}.
For the extension into the resonance region one includes
explicit resonance fields in conjunction with suitable unitarization
models or invokes  meson-exchange interactions
tested in low and intermediate energy $NN$ and $\pi N$ interactions.
At moderate energies above the prominent resonances, small-angle
$\pi\pi$ scattering falls into the domain of the Regge theory.
Large-angle scattering is expected to be dominated by pQCD
mechanisms.

There is only very limited information on
$\pi \pi$ scattering above resonances. Here the information
about the $\pi \pi$ total cross section comes from the absorption
model analysis of the experimental data on $\pi N \to X N, X \Delta$
reactions \cite{ZS84}.
There are no direct experimental data on hard $\pi\pi$ scattering.

Below I review results obtained recently in \cite{SNS01}.

\section{The pQCD two-gluon exchange}

\begin{figure}
  \begin{center}
    \includegraphics[width=5.9cm]{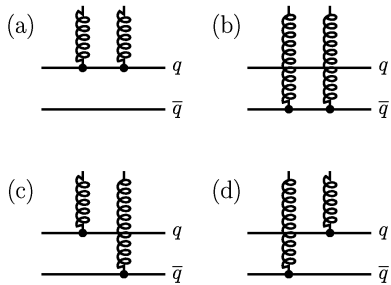}
  \end{center}
\vspace{-1.0cm}
\caption{\small
The diagrams which enter into the impact factors.}
\label{fig_gluon}
\end{figure}

Let us start with evaluation of the pQCD two-gluon contribution to
the elastic $\pi\pi$ scattering. We treat the pion as the
quark-antiquark state. The relevant pQCD diagrams which contribute
to the pion impact factor are shown in Fig.\ref{fig_gluon}.
We use both nonrelativistic and
the light-cone description of the pion \cite{SNS01}.

The impact factor representation of the pion-pion scattering
amplitude reads
\begin{eqnarray}
A(\vec{q}) = i s \cdot \frac{2}{9} \cdot \frac{1}{(2\pi)^2} \cdot
\int d^2 \kappa \; g_s^2(\kappa_1^2) g_s^2(\kappa_2^2)
\Phi_{\pi \rightarrow \pi}^{2g}(\vec{q},\vec{\kappa})
\Phi_{\pi \rightarrow \pi}^{2g}(\vec{q},\vec{\kappa})
\nonumber \\
\frac{1}{(\vec{q}/2 + \vec{\kappa})^2}
\frac{1}{(\vec{q}/2 - \vec{\kappa})^2} \;,
\label{amplitude_impactfactors}
\end{eqnarray}
where $\vec{\kappa}_{1/2} = \frac{\vec{q}}{2} \pm \vec{\kappa}$ are the
exchanged-gluon momenta, which are purely transverse,
2/9 is the QCD color factor for the $\pi\pi$ scattering
and $g_s$ is the QCD strong charge.

\vspace{-0.1cm}
\begin{figure}
  \begin{center}
\hspace{-1.4cm}
    \includegraphics[width=5.9cm]{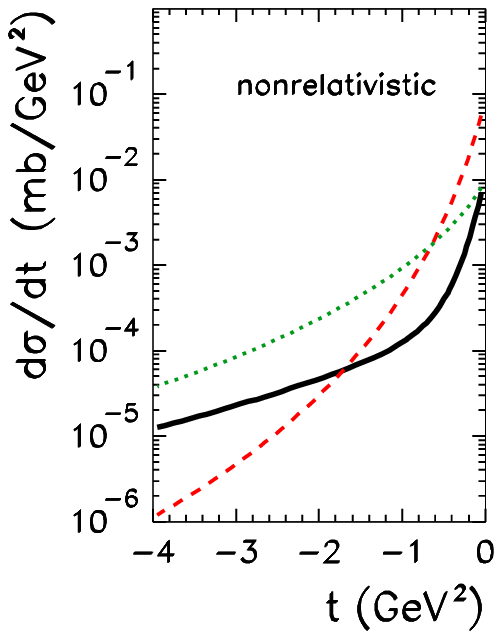}
\hspace{0.0cm}
    \includegraphics[width=5.9cm]{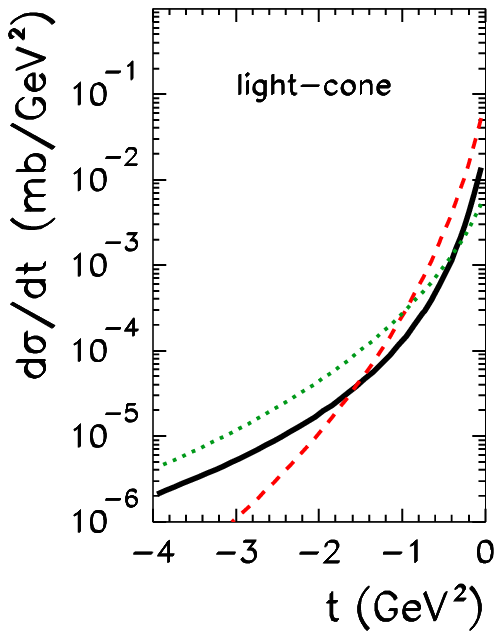}
  \end{center}
\vspace{-1.3cm}
\caption{\small
Elastic $\pi\pi$ scattering for the
2g-exchange model.
The dashed line is for pure IA contributions
(only diagrams (a) and (b) in Fig.\ref{fig_gluon}),
whereas the dotted line corresponds to the pure GGL
terms
(only diagrams (c) and (d) in Fig.\ref{fig_gluon}).
The thick solid line corresponds to the full result with
all terms for the impact factor.
}
\label{fig_pQCD_deco}
\end{figure}

At $|t| \gtrsim$ 2 GeV$^2$ the hard contribution takes
over the soft contribution to be discussed in more detail below.
The result obtained with only impulse approximation terms
(a) and (b) (dashed line) and with only Glauber-Gribov-Landshoff
\cite{Glauber,Gribov,Landshoff}
terms (c) and (d) (dotted line) are shown separately for illustration in
Fig.\ref{fig_pQCD_deco}.
The impulse approximation components of the impact factors
dominate at small to moderate values of $|t| < $ 0.5 GeV$^2$,
where the nonrelativistic (NR) and light-cone (LC) amplitudes are
nearly identical. A comparison of the NR and LC cases shows clearly
a substantial suppression of the GGL contribution by the $q-z$
correlations inherent to the LC case. The GGL
mechanism dominates at $|t| \gtrsim $ 1.0 GeV$^2$ where the LC
amplitude decreases faster than the NR one.
One should note, however, that even at large $|t|$ $\sim$ 4 GeV$^2$
due to interference effects all contributions must be included.

\section{Soft reggeon exchanges and multiple scattering}

In the case of $\pi\pi$ scattering in the considered region of
energies the soft pomeron exchange must be supplemented
by the subleading isoscalar ($f$) and isovector ($\rho$) reggeon
exchanges. For the purposes of our analysis we resort to
the simplest Regge-inspired phenomenological form:
\begin{eqnarray}
A_{\Pom}(t) &=& i \; C_{\Pom} \cdot (s/s_0)^{\alpha_{\Pom}(t)}
\cdot F_{\Pom}^2(t) \; ,
\nonumber \\
A_{f}(t)  &=& -\eta_f(t) \; C_f \cdot (s/s_0)^{\alpha_f(t)}
\cdot F_f^2(t) \; ,
\nonumber \\
A_{\rho}(t) &=& -\eta_{\rho}(t) \; C_{\rho} \cdot (s/s_0)^{\alpha_{\rho}(t)}
\cdot F_{\rho}^2(t) \; ,
\label{reggeon_amplitudes}
\end{eqnarray}
where $\eta_f$ and $\eta_{\rho}$ are somewhat simplified
signature factors \cite{SNS01}.

In the following we take the exponential
$F(t) = exp(\frac{B}{4}t)$ parametrization
for the pion-pion-reggeon vertex form factor.
For simplicity we assume one universal slope parameter for
all reggeons $B_{\Pom} = B_f = B_{\rho} \equiv B$.
In the region of small-angle scattering, in loose analogy to the
electromagnetic form factor of the pion, we consider also
the monopole $F(t) = \frac{1}{1-B_{mon}t}$ parametrizations
for the pion-pion-reggeon vertex form factor, again with
one universal parameter $B_{mon}$ \cite{SNS01}.

Assuming Regge factorization the residues at $t=0$ can be evaluated
from those for $\pi N$ and $NN$ scattering as:
\begin{equation}
C^{\pi\pi}_i = {(C^{\pi N}_i)^2 \over  C^{NN}_i}
\label{factorization}
\end{equation}
for each reggeon considered $i = \Pom, f, \rho$.
Then the corresponding Regge phenomenology of
$\pi N$ and $NN$ scattering \cite{DL92} gives
$C_{\Pom}$ = 8.56 mb, $C_f$ = 13.39 mb and $C_{\rho}$ = 16.38 mb.
Absorption corrections, generated by Regge cuts, are known to break
the factorization (\ref{factorization}).
We take for the pomeron trajectory $\alpha_{\Pom}(0)$ = 1 and
$\alpha_{\Pom}^1$ = 0.25 GeV$^{-2}$
and for both subleading trajectories $\alpha_R(0)$ = 0.5 and
$\alpha_R^1$ = 0.9 GeV$^{-2}$, i.e. values well known from
the Regge phenomenology \cite{Collins}.
In the above evaluation we have neglected the possible
small pQCD 2g-exchange contribution to the  $\pi N$ and $NN$ total
cross section, which is justified for the purpose
of our exploratory study.
Thus the slope $B$ is our basic free parameter.

The total single-reggeon exchange amplitude is now
\begin{equation}
A_{soft}^{1-st}(t) = A_{\Pom}(t) + A_f(t) + \xi A_{\rho}(t)  \; ,
\end{equation}
where $\xi$ = -1 for $\pi^+ \pi^-$, $\xi$ = 0 for $\pi^{\pm} \pi^0$
and $\xi$ = 1 for two identical pions.

\vspace{-0.1cm}
\begin{figure}
  \begin{center}
\hspace{-1.4cm}
    \includegraphics[width=5.9cm]{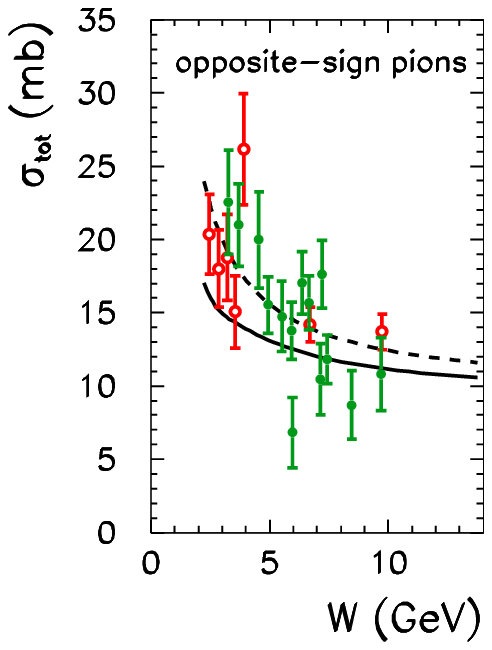}
\hspace{0.0cm}
    \includegraphics[width=5.9cm]{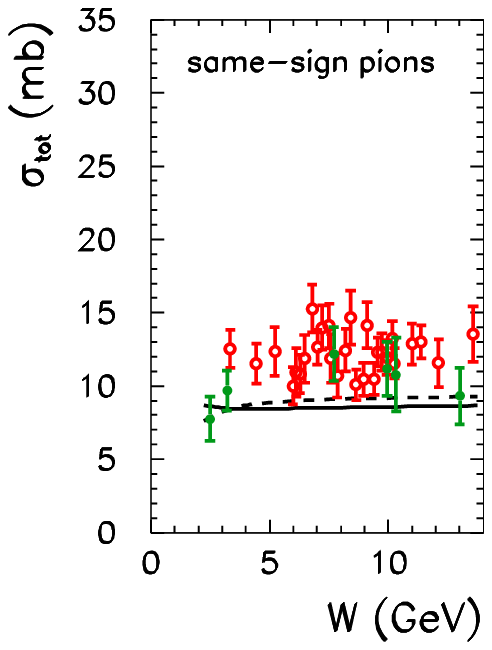}
  \end{center}
\vspace{-1.3cm}
\caption{\small
Total cross section for $\pi^+ \pi^-$ (left panel)
and $\pi^+\pi^+$ or $\pi^-\pi^-$ (right panel) scattering
as a function of center-of-mass energy $W$.
The experimental data are from \cite{ZS84}.
The single pomeron and subleading reggeon exchanges
are given by the dashed lines. The solid line is obtained from the
dashed line after including the absorption corrections.
}
\label{fig_tot}
\end{figure}

The total cross section for the same-sign and opposite-sign
$\pi\pi$ calculated with single reggeon exchanges,
including small hard two-gluon component,
is shown in Fig.\ref{fig_tot} by the dashed line.
The thick solid line includes in addition the absorption
corrections evaluated in the double-scattering approximation.
Our predictions well coincide with total cross sections extracted
in \cite{ZS84} from the absorption Regge model analysis of
$\pi N \to XN, X\Delta$ reactions.
While the opposite-sign $\pi\pi$ total cross section depends strongly
on energy, the same-sign $\pi\pi$ total cross section is almost
independent of energy.

Regge cut or absorption corrections to single reggeon+pomeron exchange
have been studied actively in the past (see for instance \cite{T-M}).
Here we restrict ourselves to the dominant double-scattering corrections
which read
\begin{equation}
A_{ij}^{(2)}(s,\vec{k}) =
\frac{i}{32 \pi^2 s} \int d^2 \vec{k}_1 d^2 \vec{k}_2 \;
\delta^2 (\vec{k} - \vec{k}_1 - \vec{k}_2) \;
A_i^{(1)}(s,\vec{k}_1) \; A_j^{(1)}(s,\vec{k}_2) \; .
\label{double_scattering}
\end{equation}
In general, the single scattering amplitudes $A_k^{(1)}$ in
(\ref{double_scattering}) are not restricted to soft reggeon
exchanges and hard two-gluon exchanges should be included too.
Consequently in the following we shall include
the (soft $\otimes$ soft), (soft $\otimes$ hard)+(hard $\otimes$ soft)
and (hard $\otimes$ hard) double-scattering amplitudes. The last three
double-scattering contributions are expected to be small, at least
at forward angles, compared to the leading (soft $\otimes$ soft)
absorption correction.

\vspace{-0.1cm}
\begin{figure}
  \begin{center}
\hspace{-1.4cm}
    \includegraphics[width=5.9cm]{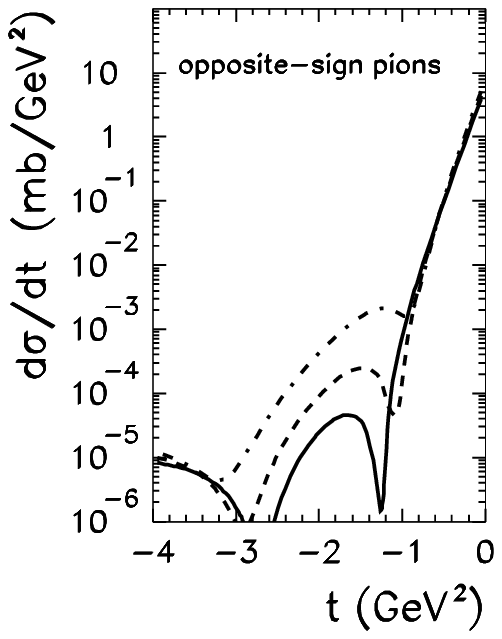}
\hspace{0.0cm}
    \includegraphics[width=5.9cm]{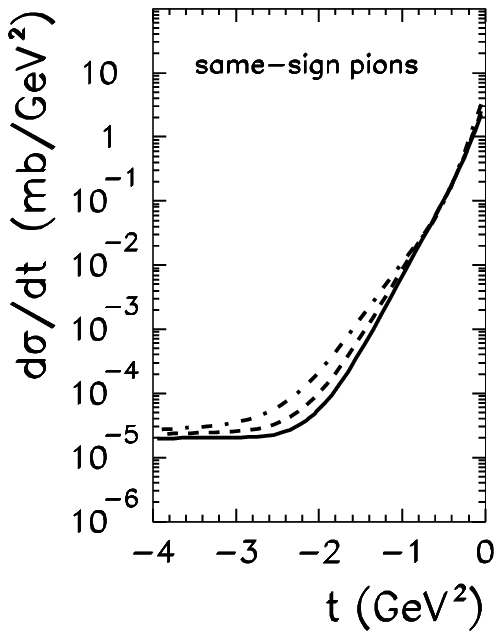}
  \end{center}
\vspace{-1.3cm}
\caption{\small
The $t$-dependence of the $\pi\pi$ elastic cross section
with the inclusion of all single- and double-exchange contributions
for different center of mass energies
3 (dash-dotted), 4 (dashed), 5 (solid) GeV.
}
\label{fig_diff_energies}
\end{figure}

Let us concentrate now on angular distributions.
It is interesting how the diffractive pattern of the angular distribution
changes with energy.
In Fig.\ref{fig_diff_energies} we present angular distributions
for $\pi^+\pi^-$ (left panel) and the same-sign pion-pion (right
panel) elastic scattering across our
region of interest for W =  3, 4, 5 GeV for exponential
vertex form factor with $B$ = 4 GeV$^{-2}$.
The small-$t$ differences at different energies come predominantly
from the energy dependence of the subleading reggeons.
The difference in the region of intermediate $t$
is due to interference of single- and double-scattering terms.
Only at very large $t \sim$ 4 GeV$^2$ the cross section starts to
approximately scale with energy.

\section{Conclusions}

We have investigated the
$\pi\pi$ scattering in the region of intermediate energies W = 2 - 5 GeV.

First, we have investigated 2g-exchange mechanism in the elastic
$\pi\pi$ scattering within a relativistic approach to the pion wave
function.
We have found the dominance of the impulse approximation terms at
small $|t|$ and the Glauber-Gribov-Landshoff terms at large $|t|$.

Assuming dominance of soft physics and Regge factorization at small
$|t|$ we have predicted the total cross section for $\pi\pi$
scattering consistent with experimental values extracted in
the literature.
The interplay of soft and hard processes in multiple scattering was
analysed. We have found
strong interference effects between single soft and hard amplitudes
and some of dominant double-scattering amplitudes.
We predict rather different angular distributions of elastic
scattering for opposite-sign pions and for same-sign pions.


\end{document}